\documentclass[twocolumn,showpacs,preprintnumbers,draftclsnofoot]{revtex4}
\usepackage{amsmath}
\usepackage{amssymb}
\usepackage{graphicx}
\usepackage{dcolumn}
\usepackage{bm}
\usepackage{amssymb}
\usepackage{graphicx}
\usepackage{dcolumn}
\usepackage{bm}

\begin{document}

\title{ Spin Imbalanced anomalous Hall effect and Magnus Hall effect of Bilayer Silicene  }
\author{Ma Luo\footnote{Corresponding author:luom28@mail.sysu.edu.cn} }
\affiliation{The State Key Laboratory of Optoelectronic Materials and Technologies \\
School of Physics\\
Sun Yat-Sen University, Guangzhou, 510275, P.R. China}

\begin{abstract}

Silicene have been proposed to host multiple topological phases depending on the environmental parameters such as gated voltage and proximity induced exchange field. Two typical topological phase are quantum spin Hall phase and quantum anomalous Hall phase. We found that in the presence of antiferromagnetic exchange field and (or) gated voltage, the bilayer silicene host a topological phase that the Chern numbers of the two spin are different in both sign and magnitude, .i.e the anomalous Hall effects of the two spins are imbalanced. The topological phase is generated by triple band inversion that changes the Chern number of the upper and lower bands by $\pm3$. The phase host large anomalous spin Hall respondence. Meanwhile, because of the large trigonal warping in bilayer silicene near to the triple band inversion regime, the Magnus Hall effect is large. This finding enriches the topological systems of bilayer silicene as candidate for spintronic devices.

\end{abstract}

 \maketitle

\section{Introduction}

Silicene is monolayer of silicon atoms in buckle honeycomb lattice, which exhibit varying topological phase due to the sizable spin-orbit coupling (SOC) \cite{YuguiYao2011,YuguiYao2011a,Alessandro17}. By tuning the gated voltage and proximity induced magnetic exchange field, the topological phases of the silicene is switched among being quantum anomalous Hall (QAH), quantum spin Hall (QSH), spin-polarized QAH (SQAH), and quantum spin每quantum anomolous Hall (QSQAH), valley-polarized metal (VPM), and multiple trivial insulating phases \cite{Motohiko12,Motohiko13,Motohiko13a,Motohiko13b,Motohiko14}. The tight binding model of monolayer silicene is similar to that of monolayer graphene. However, the situation change for bilayer systems. For bilayer graphene, the two monolayer graphenes are bonding by weak Van der Waals force \cite{Avouris07,CastroNeto09}. By contrast, bilayer silicene is form by AB stacking of two monolayer silicene with strong inter-layer chemical bonding \cite{YuguiYao2013,YuguiYao2015}. In the tight binding model, the inter-layer hopping terms are larger than the intra-layer hopping term. As a result, the electronic and magnetic structure as well as the topological properties of bilayer silicene is expected to be different from that of bilayer graphene. Recent theoretical studies showed that the bilayer silicene could become superconductor \cite{YuguiYao2013,YuguiYao2015}, or host spin polarized current \cite{ZhangYuzhong2018}.

One of the motivation of studying silicene is to design spin valve for spintronic application \cite{Pesin12,WeiHan14,Kheirabadi14,YangWang15}. As a result, it is important to master the topological properties of the bilayer silicene under varying physical parameters. In this paper, we study the topological phase of bilayer silicene with the presence of gated voltage, antiferromagnetical exchange field and the intrinsic SOC. Because of the strong interlayer hopping in bilayer silicene, within a large phase regime of the gated voltage and exchange field, the band gap is not opened while the direct band gap at all $\mathbf{k}$ points throughout the whole Brillouin zone remain open. Thus, the Chern number of the valance and conduction band can be well defined, which is the integration of the Berry curvature through the whole Brillouin zone \cite{Thouless1982,Kohmoto1985,ZhenhuaQiao10}. Within a regime of the phase diagram, the Chern number of one spin is $\pm3$ and that of the other spin is $\mp2$, so that the QAH effects of two spin are imbalanced. The band structure of nanoribbon shows that the robust edge states appear across the direct bulk band gap. At the phase transition point between Chern number being $\pm3$ and $0$, three band crossings simultaneously occur at three $\mathbf{k}$ points beyond the $K$ (or $K^{\prime}$) point along the three $K-\Gamma$ (or $K^{\prime}-\Gamma$) directions, which is designated as triple band inversion. The three $\mathbf{k}$ points are designated as $K_{(3-\Gamma)}$ (or $K^{\prime}_{(3-\Gamma)}$). In the phase regime near to the triple band inversion, the Berry curvature near to $K^{(\prime)}_{(3-\Gamma)}$ is large, which generate large Hall respondence. The spin Hall conductance \cite{JairoSinova15}, anomalous Hall conductance and Magnus Hall conductance \cite{FuLiang2019} are calculated. In the phase that break time reversal symmetric, the anomalous Hall conductance is sizable. In the phase that the time reversal symmetric is preserved, the anomalous Hall conductance is absence, while the Magnus Hall conductance is sizable if the system is in the ballistic limit \cite{FuLiang2019}.

This parer is organized as follow: In section II, the tight binding model of the bilayer silicene is reviewed. In section III, the bulk band structure and topological phase diagram is discussed. In section IV, the band structure of nanoribbon is discussed. In section V, the spin Hall conductance, anomalous Hall conductance and Magnus Hall conductance are calculated and discussed. In section VI, the conclusion is given.

\section{Tight Binding Model of Bilayer Silicene}

The atomic structure of the bilayer silicene is similar to the AB stacking bilayer graphene, except that the A and B sublattice in the same layer have different z coordinate. We designate the four atomic site in one unit cell as $|A,t\rangle$, $|B,t\rangle$, $|A,b\rangle$, and $|B,b\rangle$. $|B,t\rangle$ is on top of $|A,b\rangle$ with interlayer distance being $l_{v}=2.53$ ${\AA}$. The z coordinate of $|A,t\rangle$ ($|B,b\rangle$) is $l_{z}=0.46$ ${\AA}$ higher (lower) than that of $|B,t\rangle$ ($|A,b\rangle$). The tight binding Hamiltonian is given as \cite{YuguiYao2013,YuguiYao2015}
\begin{eqnarray}
H&=&t_{n}\sum_{\langle i,j\rangle,\alpha}{c^{\dag}_{i,\alpha}c_{j,\alpha}}+\frac{i\lambda_{KM}}{3\sqrt{3}}\sum_{\langle\langle i,j\rangle\rangle,\alpha,\alpha'}{\nu_{i,j}\hat{s}^{z}_{\alpha,\alpha'}c^{\dag}_{i,\alpha}c_{j,\alpha'}} \nonumber \\
&+&t_{1}\sum_{\langle i,j\rangle_{\langle A,b|B,t\rangle},\alpha}{c^{\dag}_{i,\alpha}c_{j,\alpha}}+t_{3}\sum_{\langle i,j\rangle_{\langle B,b|A,t\rangle},\alpha}{c^{\dag}_{i,\alpha}c_{j,\alpha}}\nonumber \\ &+&t_{2}\sum_{\langle i,j\rangle_{\langle A,b|A,t\rangle},\alpha}{c^{\dag}_{i,\alpha}c_{j,\alpha}}+t_{2}\sum_{\langle i,j\rangle_{\langle B,b|B,t\rangle},\alpha}{c^{\dag}_{i,\alpha}c_{j,\alpha}}\nonumber \\
&+&\sum_{i_{|A,t\rangle},\alpha}{(V+M\hat{s}^{z}_{\alpha,\alpha}+\Delta_{I})c^{\dag}_{i,\alpha}c_{i,\alpha}}\nonumber \\
&+&\sum_{i_{|B,t\rangle},\alpha}{(fV+fM\hat{s}^{z}_{\alpha,\alpha})c^{\dag}_{i,\alpha}c_{i,\alpha}}\nonumber \\
&+&\sum_{i_{|A,b\rangle},\alpha}{(-fV-fM\hat{s}^{z}_{\alpha,\alpha})c^{\dag}_{i,\alpha}c_{i,\alpha}}\nonumber \\
&+&\sum_{i_{|B,b\rangle},\alpha}{(-V-M\hat{s}^{z}_{\alpha,\alpha}+\Delta_{I})c^{\dag}_{i,\alpha}c_{i,\alpha}}
\end{eqnarray}
, where $i$($j$) is the index of the atomic site, $\alpha=\pm1$ represent spin up and down. The first term is the intra-layer nearest neighbor hopping with strength $t_{n}=1.130$ eV. The second term is intra-layer the next nearest neighbor SOC term given by the Kane-Mele model with SOC strength being $\lambda_{KM}$,  $\nu_{i,j}=\pm1$ for right or left turning hopping, and $\hat{s}^{z}$ being the Pauli matrix of spin-z. The third term is the nearest neighbor inter-layer hopping between $|B,t\rangle$ and $|A,b\rangle$ sublattice with strength being $t_{1}=2.025$ eV. The fourth and fifth terms are the next nearest neighbor inter-layer hopping with strength being $t_{2}=-0.152$ eV. The sixth term is the third nearest neighbor inter-layer hopping with strength being $t_{e}=0.616$ eV. The remaining terms are the on-site potential due to gate voltage $V$, proximity induced exchange field $M$, and effective on-site potential at $|A,t\rangle$ ($|B,b\rangle$) atomic sites $\Delta_{I}=-0.069$ eV. The exchange fields of top and bottom layer are opposite to each other. Because the z coordinate of $|A,t\rangle$ ($|B,b\rangle$) and $|B,t\rangle$ ($|A,b\rangle$) atomic sites are different, the potential at $|B,t\rangle$ ($|A,b\rangle$) atomic sites have a factor $f=(l_{v}-2l_{z})/l_{v}$. The bulk and zigzag nanoribbon band structures are obtained by diagonalization of the Hamiltonian with Bloch periodic boundary condition.

\section{Bulk Band Structure and Topological Phase Diagram}

\begin{figure}[tbp]
\scalebox{0.33}{\includegraphics{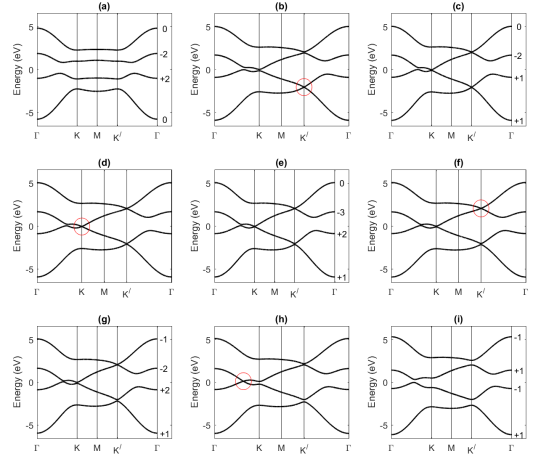}}
\caption{ The band structures of spin up electrons with $\lambda_{KM}=0.2\times3\sqrt{3}$ eV and $V+M=0$ eV in (a), 0.4807 eV in (b), 0.5 eV in (c), 0.51962 eV in (d), 0.531 eV in (e), 0.5423 eV in (f), 0.55 eV in (g),  0.60083 eV in (h), 0.8 eV in (i). In figures (a,c,e,g,i), all bands are gapped from each other and the Chern number of each band, $\bar{C}_{i}$, is labeled to the right of the band structure. In figures (b,d,f,h), the band crossings with closed gap are marked by the red circle.  }
\label{fig_bulk}
\end{figure}

For realistic silicene, the strength of the SOC is $\lambda_{KM}=0.0039\times3\sqrt{3}$ eV. The strength could be enhanced by adatom doping. In order to studied the quality properties of the topological phase diagram, we consider the model with larger SOC in this section, assuming $\lambda_{KM}=0.2\times3\sqrt{3}$ eV. The band structures of spin up electrons with varying strength of the sum of gated voltage and exchange field $(V+M)$ are plotted in Fig. \ref{fig_bulk}. The Chern number of each band, which is designated as $\bar{C}_{i}$ with $i$ being the band index, can be calculated by integrating the Berry curvature throughout the whole Brillouin zone. The total Chern number of one spin $C_{\alpha}$ is the summation of $\bar{C}_{i}$ of the two valence bands, and the total Chern number of the system is $C=C_{+}+C_{-}$. With $V+M=0$ eV, the spin up band is Chern insulator with $C_{+}=+$2. As $(V+M)$ increases, gap closing occur at four critical value of $(V+M)$ as shown in Fig. \ref{fig_bulk} (b,d,f,h). The gap closing at the three critical value with $V+M=0.4807$, $0.51962$, $0.5423$ eV in (b), (d), (f) occur between the first and second, second and third, third and fourth band at $K^{\prime}$, $K$, $K^{\prime}$ points, respectively. After the gap reopen, the Chern number of the band above (below) the band crossing point decrease (increase) one. The gap closing at the fourth critical value with $V+M=0.60083$ eV occur between the lower conduction band and the higher valence band at $K_{(3-\Gamma)}$, as shown in Fig. \ref{fig_bulk} (h). The Chern number of the lower conduction band (higher valence band) decrease (increase) three, because there are simultaneously three band crossing.


\begin{figure}[tbp]
\scalebox{0.5}{\includegraphics{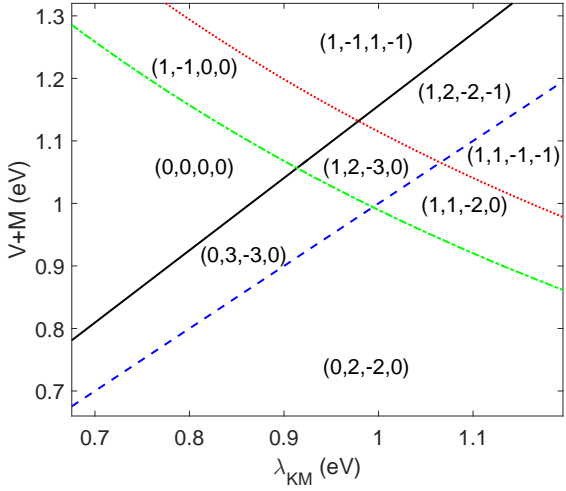}}
\caption{ The phase diagram of the Chern number of each band in the $(V+M)-\lambda_{KM}$ space. The text within each regime is $(\bar{C}_{1},\bar{C}_{2},\bar{C}_{3},\bar{C}_{4})$. This phase diagram is for the spin up electron. }
\label{fig_phase2}
\end{figure}

\begin{figure}[tbp]
\scalebox{0.33}{\includegraphics{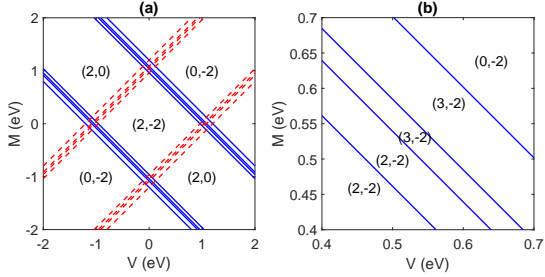}}
\caption{ The phase diagram of the Chern number of spin up and spin down band in the $V-M$ space with $\lambda_{KM}=0.2\times3\sqrt{3}$ eV. The text within each regime is $(C_{+},C_{-})$. The solid blue and dashed red lines are the phase boundary of the Chern number of spin up and down electron, respectively. }
\label{fig_phase}
\end{figure}

The critical value of the gap closing points are dependent on $\lambda_{KM}$. For the gap closing points at $K$ or $K^{\prime}$ points, the critical value of $V+M$ can be obtained by solving the eigenvalue problem of the Hamiltonian. At these $\mathbf{k}$ points, the Hamiltonian become
\begin{equation}
H=\left[\begin{array}{cccc}
H_{11} & 0 & 0 & 0 \\
0 & H_{22} & t_{1} & 0 \\
0 & t_{1} & H_{33} & 0 \\
0 & 0 & 0 & H_{44} \\
\end{array}\right]\label{HamiltonianAtK}
\end{equation}
, where $H_{11}=V+M+\Delta_{I}+\lambda_{KM}\eta$, $H_{22}=f(V+M)-\lambda_{KM}\eta$, $H_{33}=-f(V+M)+\lambda_{KM}\eta$, $H_{44}=-V-M+\Delta_{I}-\lambda_{KM}\eta$, with $\eta=\pm1$ for $K$ and $K^{\prime}$ valley. The band crossing between the lower conduction band and the higher valence band occur at $V+M=\lambda_{KM}$, which is plotted as dashed blue line in Fig. \ref{fig_phase2}. The band crossing between the two valence (conduction) band occur at
\begin{widetext}
\begin{equation} V+M=\frac{\pm\Delta_{I}-(1+f)\lambda_{KM}+\sqrt{[\mp2\Delta_{I}+2(1+f)\lambda_{KM}]^2-4(1-f^2)(\Delta_{I}^2\mp2\Delta_{I}\lambda_{KM}-t_{1}^{2})}/2}{1-f^{2}}
\end{equation}
\end{widetext}
, which is plotted as dashed-dotted green (dotted red) line in Fig. \ref{fig_phase2}. The critical value of the triple band inversion is calculated numerically, and plotted as solid black line in Fig. \ref{fig_phase2}. The phase diagram for spin down electron can be obtained by replacing $V+M$ with $V-M$.

Assuming $\lambda_{KM}=0.2\times3\sqrt{3}$ eV, the phase diagram in the $V-M$ space for both spin is plotted in Fig. \ref{fig_phase}. In some regimes, the Chern number of one spin is zero, and the Chern number of the other spin is $\pm2$. In some other regimes, the Chern number of both spin is nonzero with opposite sign but different magnitude. These phase regimes are designated as spin imbalanced quantum anomalous Hall phase.

\section{Band Structure of Zigzag Nanoribbon}

\begin{figure}[tbp]
\scalebox{0.52}{\includegraphics{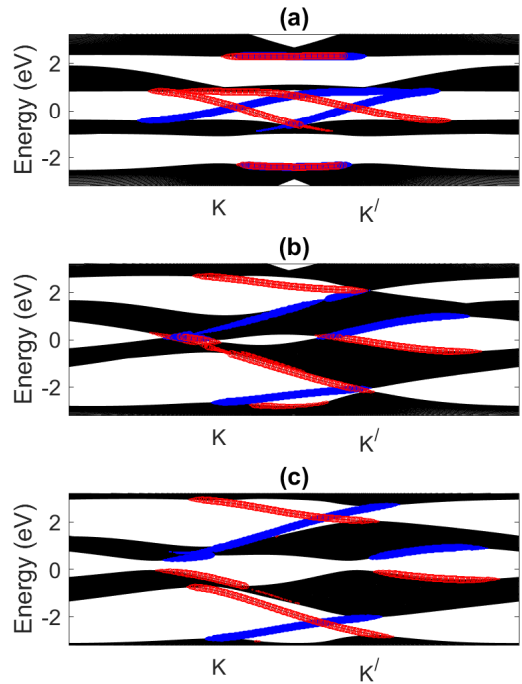}}
\caption{ The band structure of the zigzag nanoribbon with 60 unit cells at the width direction and $V=M=0$ eV in (a), 0.55 eV in (b), 0.85 eV in (c). Only spin up electron is plotted. The solid black lines are the band structures; the size of the filled blue and empty red circular marker represents the weight of the wave function within two unit cells beyond the open boundary at the left and right zigzag edge, respectively.  }
\label{fig_ribbon1}
\end{figure}

In order to visualized the bulk-boundary correspondence, the band structure of the zigzag nanoribbon with the bulk in varying topological phase are studied. Assuming $\lambda_{KM}=0.2\times3\sqrt{3}$ eV, the band structure of the spin up electron of the zigzag nanoribbon with 60 unit cells at the width direction and $V=M=0$, 0.55, and 0.85 eV are plotted in Fig. \ref{fig_ribbon1}(a), (b) and (c), respectively. In Fig. \ref{fig_ribbon1}(a), the bulk state have Chern number being $C_{+}=2$, so that there are two pairs of edge state across the zero energy. In Fig. \ref{fig_ribbon1}(b), the bulk state is corresponding to that in Fig. \ref{fig_bulk}(g). The global indirect band gap of bulk is close, but the direct band gap throughout the whole Brillouin zone is nonzero. Within the direct band gap, the summation of the Chern numbers of the lower (higher) two bands is +3 (-3), so that there are three pairs of edge bands across this gap. Within the band gap between two valence (conduction) bands, there is one pairs of edge bands, because the summation of the Chern number of the lower and higher band are $\pm1$. In Fig. \ref{fig_ribbon1}(c), the Chern number of the bulk is zero, so that there is no edge state across the zero energy. The Chern numbers of the two valence (conduction) bands are $\pm1$ ($\mp1$), so that there is one pair of edge bands between the two valence (conduction) bands.

\begin{figure}[tbp]
\scalebox{0.31}{\includegraphics{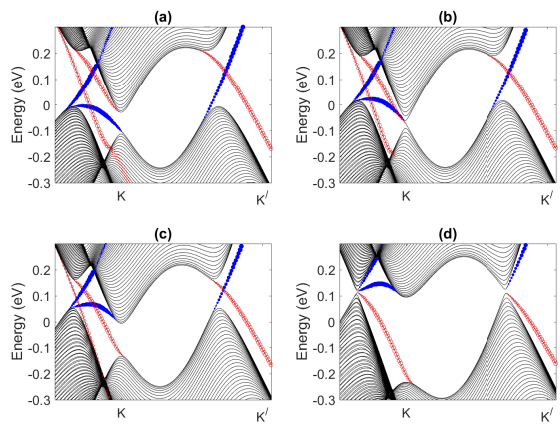}}
\caption{ The band structure of the zigzag nanoribbon with 60 unit cells at the width direction and $V=M=0.5$ eV in (a), 0.51962 eV in (b), 0.55 eV in (c), 0.60083 eV in (d). Only spin up electron is plotted. The solid black lines are the band structures; the size of the filled blue and empty red circular marker represents the weight of the wave function within two unit cells beyond the open boundary at the left and right zigzag edge, respectively. }
\label{fig_ribbon2}
\end{figure}

In order to further visualized the triple band inversion that result in change of Chern number by three, the band structure of four zigzag nanoribbon are given in Fig. \ref{fig_ribbon2}. For the nanoribbon in Fig. \ref{fig_ribbon2}(b) and (c), the direct band gap of bulk close at the $K$ point and $K_{(3-\Gamma)}$, respectively. In Fig. \ref{fig_ribbon2}(a), before the band inversion at $K$ point, there are two pairs of edge bands across the bulk gap, and one pairs of edge bands that does not cross the bulk gap. Once the bulk gap close at the $K$ point, the pair of edge bands inverse at the $K$ point. When the gap reopen, as shown in Fig. \ref{fig_ribbon2}(c), there are three pairs of edge bands across the bulk gap. At the critical point in Fig. \ref{fig_ribbon2}(d), triple band inversion occur at $K_{(3-\Gamma)}$ in bulk. In zigzag nanoribbon, two (one) of the three $\mathbf{k}$ points in $K_{(3-\Gamma)}$ are projected to the $k$ point to the left (right) of the $K$ point of zigzag nanoribbon, where the two (one) pairs of the edge bands inversion occur. After the gap reopen, the three pairs of edge bands are gapped simultaneously. Similar type of edge state was found in $\mathbb{Z}_{2}$ topological semi-metal in bilayer systems \cite{HuiPan14}.

\section{Hall respondences}

Although the numerical result in the previous section can visualize the edge band, the experimental measurement of the exotic edge band is non-realistic because the SOC is chosen to be very large. In this section, we use a realistic SOC in bilayer silicene with $\lambda_{KM}=0.0039\times3\sqrt{3}$ eV, and calculate the spin Hall conductance, anomalous Hall conductance and Magnus Hall conductance. Two types of systems are studied: (i) the system that break time reversal symmetric and have imbalance Chern number of the two spins, i.e. $V=M=0.0111$ eV with $C_{+}=3$ and $C_{-}=-2$; (ii) the system that preserve time reversal symmetric and near to the triple band inversion point, i.e. $M=0$ and $V=0.0111$, $0.0222$ or $0.0333$ eV with $(C_{+},C_{-})=(+2,-2)$, $(+3,-3)$ or $(0,0)$, respectively.

\subsection{Break time reversal symmetric}

\begin{figure}[tbp]
\scalebox{0.58}{\includegraphics{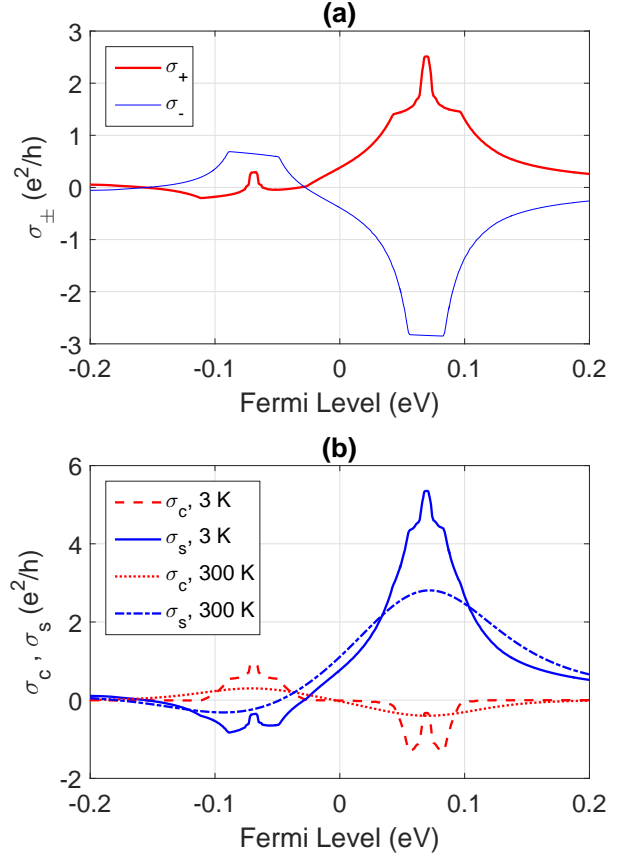}}
\caption{ (a) The Hall respondence of spin up $\sigma_{+}$ and spin down $\sigma_{-}$ versus the Fermi level are plotted as thick red and thin blue lines with the temperature being 3 K, respectively. (b) The spin Hall conductance $\sigma_{s}$ and anomalous Hall conductance $\sigma_{c}$, versus the Fermi level is plotted as solid blue (dashed-dotted blue) and dashed red (dotted red) line with the temperature being 3 K (300 K), respectively. The bulk bilayer silicene with realistic SOC being $\lambda_{KM}=0.0039\times3\sqrt{3}$ eV, and $V=M=0.0111$ eV is applied. }
\label{fig_HallSigma}
\end{figure}

Assuming $V=M=0.0111$ eV, the system is within the regime between the solid black and dashed blue line in Fig. \ref{fig_phase2}) with $C_{+}=3$ and $C_{-}=-2$, and the directive band gap at all $\mathbf{k}$ points throughout the whole Brillouin zone is maximized. If the global band gap is opened, the anomalous Hall conductance and spin Hall conductance is quantized as $(C_{+}+C_{-})e^{2}/h=e^{2}/h$ and $(C_{+}-C_{-})e^{2}/h=5e^{2}/h$, respectively. However, the bulk band of the spin up electron is indirectly closed. The bulk band interfere with the Hall respondence. Thus, the Hall conductance is dependent on the Fermi level, as shown in Fig. \ref{fig_HallSigma}. The Hall conductance of spin $\alpha$ is calculated by the equation \cite{JairoSinova15}
\begin{equation}
\sigma_{\alpha}=\frac{e^2}{h}\int{\frac{d^{2}\mathbf{k}}{(2\pi)^2}\sum_{n}{f^{\alpha}_{n\mathbf{k}}\Omega^{\alpha}_{n,xy}(\mathbf{k})}}
\end{equation}
\begin{equation}
\Omega^{\alpha}_{n,xy}(\mathbf{k})=-2Im\sum_{n'\ne n}{\frac{\langle n\mathbf{k},\alpha|\hat{v}_{x}|n'\mathbf{k},\alpha\rangle\langle n'\mathbf{k},\alpha|\hat{v}_{y}|n\mathbf{k},\alpha\rangle}{(E^{\alpha}_{n\mathbf{k}}-E^{\alpha}_{n'\mathbf{k}})^{2}}}
\end{equation}
,where $f^{\alpha}_{n\mathbf{k}}$ is the Fermi-Dirac distribution of the $\alpha$ spin band, $\hat{v}_{x(y)}$ is the velocity operator. The anomalous Hall conductance and spin Hall conductance is given as $\sigma_{c}=\sigma_{+}+\sigma_{-}$ and $\sigma_{s}=\sigma_{+}-\sigma_{-}$, respectively. Note that the definition of the spin Hall conductance is normalized by $\hbar/e$, so that the unit is the same as anomalous Hall conductance. For the bilayer silicene with realistic SOC being $\lambda_{KM}=0.0039\times3\sqrt{3}$ eV, and $V=M=0.0111$ eV, the direct band gap of spin up (down) electron at $K_{(3-\Gamma)}$ ($K^{\prime}_{(3-\Gamma)}$) is at the energy range $(0.0677,0.0709)$ eV [$(0.0551,0.0836)$ eV]. The Hall conductance $\sigma_{+}$ ($\sigma_{-}$) within the energy range have a plateau below $3e^{2}/h$ ($-2e^{2}/h$), as shown in Fig. \ref{fig_HallSigma}(a). The deviation from the quantized conductances is due to the interference from the bulk states. If the strength of the SOC is larger, the interference is weaker, so that the plateau is closer to the quantized conductances. The combination of the Hall conductance of the two spin gives the anomalous Hall conductance and spin Hall conductance, which have plateau and double plateau, respectively, as shown in Fig. \ref{fig_HallSigma}(b). If the temperature is raised to room temperature, more bulk states outside of the gap at $K_{(3-\Gamma)}$ ($K^{\prime}_{(3-\Gamma)}$) interfere with the Hall respondence, so that the plateau and double plateau for the anomalous Hall conductance and spin Hall conductance disappear.

\subsection{Preserve time reversal symmetric}

\begin{figure}[tbp]
\scalebox{0.58}{\includegraphics{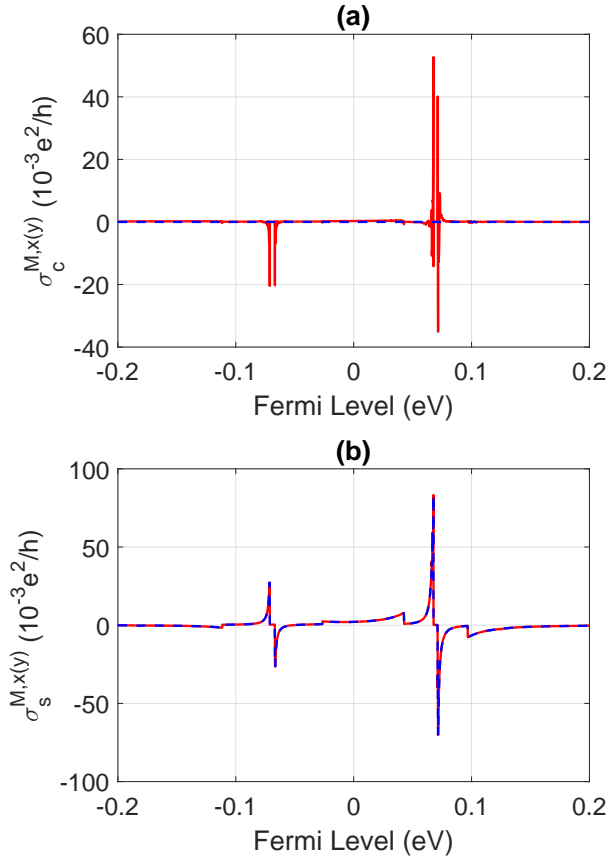}}
\caption{ (a) The Magnus charge Hall conductance $\sigma_{c}^{M,x(y)}$ versus the Fermi level. (b) The Magnus spin Hall conductance $\sigma_{s}^{M,x(y)}$ versus the Fermi level. Realistic SOC with $\lambda_{KM}=0.0039\times3\sqrt{3}$ eV and $M=0$, $V=0.0222$ eV is assumed. The Hall respondence with longitudinal bias $\Delta U=0.001$ eV along x(y) direction is plotted as blue dashed (red solid) line. }
\label{fig_MagnusHallSigma}
\end{figure}

If time reversal symmetric is preserved, i.e. $M=0$, the anomalous Hall conductance is zero. However, within the ballistic limit, the Magnus Hall effect induces Hall respondence of charge current because of the effective magnetic field due to the Berry curvature. The condition for the appearance of the Magnus Hall effect is to break the inversion symmetric by the gated voltage. From the band structure of the bilayer silicene, the trigonal warping is large. Near to the triple band inversion point, the Berry curvature near to $K_{(3-\Gamma)}$ is large. Thus, we can expect large Magnus Hall conductance. The Magnus Hall conductance of spin $\alpha$ with longitudinal bias $\Delta U$ along x(y) direction can be calculated by the equation \cite{FuLiang2019}
\begin{equation}
\sigma_{\alpha}^{M,x(y)}=\frac{e^2}{h}\frac{\Delta U}{2\pi}\int_{v_{x(y)}(\mathbf{k})>0}{d^{2}k \sum_{n}{\Omega^{\alpha}_{n,xy}(\mathbf{k})\delta(E^{\alpha}_{n\mathbf{k}}-\mu)}}
\end{equation}
with $\mu$ being the Fermi level. The Magnus charge Hall conductance and Magnus spin Hall conductance of the bilayer silicene are defined as $\sigma_{c}^{M,x(y)}=\sigma_{+}^{M,x(y)}+\sigma_{-}^{M,x(y)}$ and $\sigma_{s}^{M,x(y)}=\sigma_{+}^{M,x(y)}-\sigma_{-}^{M,x(y)}$, which is plotted in Fig. \ref{fig_MagnusHallSigma}(a) and (b), respectively. Because of the mirror symmetric along the x direction ($\Gamma-M$ direction), the Magnus charge Hall conductance with longitudinal bias along x direction is zero. The Magnus charge Hall conductance with longitudinal bias along y direction ($\Gamma-K^{(\prime)}$ direction) is sizable. With the Fermi level near to the band edge of the gap opening near to $K^{(\prime)}_{(3-\Gamma)}$, the Magnus charge Hall conductance is as large as $50\frac{e^{2}}{h}\frac{\Delta U}{1 meV}$. In contrast, the Magnus spin Hall conductance is independent on the direction of the longitudinal bias.

\section{conclusion}

In conclusion, the topological phase of bilayer silicene with intrinsic SOC, gated voltage and antiferromagnetic configured exchange field are studied. The phase regime with imbalance Chern number between spin up and spin down electron is identified. Within this phase regime, the Chern number of one spin equates to $\pm3$, which is due to the triple band inversion at $K^{(\prime)}_{(3-\Gamma)}$. For the bilayer silicene with realistic SOC, the spin Hall conductance, anomalous Hall conductance and Magnus Hall conductance are dependent on the gated voltage, exchange field as well as the Fermi level. For the phase with broken time reversal symmetric and imbalanced Chern number of the two spin, the spin Hall conductance has a plateau above $5e^{2}/h$, if the Fermi level is within the gap at $K_{(3-\Gamma)}$; the anomalous Hall conductance has double plateaus below $-1e^{2}/h$, if the Fermi level is within the gap at $K^{\prime}_{(3-\Gamma)}$. For the phase with time reversal symmetric, the Magnus Hall effect is found to be large at the band edge near to $K^{(\prime)}_{(3-\Gamma)}$. The Magnus Hall conductance is strongly dependent on the direction of the longitudinal bias.


\begin{acknowledgments}
The project is supported by the National Natural Science Foundation of China (Grant:
11704419).
\end{acknowledgments}

\clearpage

\end{document}